# Performance of Android Forensics Data Recovery Tools

Bernard Chukwuemeka Ogazi-Onyemaechi[1], Ali Dehghantanha[1]; Kim-Kwang Raymond Choo[2]

[1]School of Computing, Science and Engineering, University of Salford, Manchester, United Kingdom
[2] Information Assurance Research Group, University of South Australia, Australia
b.c.ogazi-onyemaechi@edu.salford.ac.uk; A.Dehghantanha@salford.ac.uk; Raymond.Choo@fulbrightmail.org

Abstract-
Recovering deleted or hidden data is among most important duties of forensics investigators. Extensive utilisation of smartphones as subject, objects or tools of crime made them an important part of residual forensics. This chapter investigates the effectiveness of mobile forensic data recovery tools in recovering evidences from a Samsung Galaxy S2 i9100 Android phone. We seek to determine the amount of data that could be recovered using Phone image carver, Access data FTK, Foremost, Diskdigger, and Recover My File forensic tools. The findings reflected the difference between recovery capacities of studied tools showing their suitability in their specialised contexts only.

## 1.0 INTRODUCTION

Smart mobile devices, particularly smartphones, are increasingly popular in today's Internet-connected society (1–4). For example, few years ago in 2010, shipments of smartphone grew by 74 percent to 295 million units (3,4). Unsurprisingly, sales of smartphones have been increasing since then (5,6), and it has been estimated that 1.5 billion smartphones will be sold by 2017 and 1 billion mobile subscribers by 2022 (7–15).

Such devices are generally used to make phone calls, send SMS messages, web browsing, locate places of interests, map navigation, image and video capture, entertainment (e.g. gaming and lifestyle), business and economic transactions (e.g. internet banking), take notes, create and view documents, etc. (6,16–18). Due to their widespread adoption in corporate businesses, these devices are a rich source of information (e.g. corporate data and intellectual property) (19–21) (19,22,23). The potential to target such devices for criminal activities (e.g. malware such as banking Trojans) or be used as an attack launch pad (e.g. used to gain unauthorized access to corporate data) (19,24,25) (26–28) (29,30), makes it important to ensure that we have the capability to conduct a thorough investigation of such devices (23,31) (32) (18,33–36).

While there are a (small) number of forensic tools that can be used in the forensic investigation of smart mobile devices (37), the extent to which data can be recovered varies, particularly given the wide range of mobile devices and the constant evolution of mobile operating systems and hardware (38,39). For example, recovering data from the internal memory of a smartphone remains a challenge (35,40,41). Further to these challenges is the requirement to create forensically sound and effective tools and procedures (37) (20,21,42).

Therefore, it is essential that the forensic community keeps pace with forensic solutions for smart mobile devices (43,44) (2–4). This is the focus of this chapter. Specifically, we study the effectiveness of five popular mobile forensics tools, namely: Phone Image Carver, AccessData FTK, Foremost, Recover My Files, and DiskDigger, in recovering evidential data from a factory-restored Samsung Galaxy Note 3 running Android Jelly Bean version 4.3.

The structure of this chapter is as follows. Section 2 reviews related work. Section 3 outlines the methodology and our experiment setup. Section 4 presents our findings, and Section 5 concludes this chapter.

## 2.0 RELATED WORK

The present investigation is conducted based on the current acquisition method on Smartphone and the extent of available forensic techniques and tools on the analysis of evidence. Smartphone has many profound sections of where and how evidences are collected when it is dealing with crime occurrence (45). Therefore, with the ever-increasing features and utility, it becomes much more complicating to collect evidences from a Smartphone (44,46). Consequently, there are different acquisition methods on different architectures and software that a Smartphone operates on (18,40). These differences in the architecture make it extremely difficult to perform similar acquisition method on different devices and operating systems (24). However, NIST created guidelines for Computer Forensics Tools Testing (CFTT) to provide for the differences in architecture (46). Therefore, Mobile device forensics has been defined as the science of recovering digital evidence from a mobile device under forensically sound conditions using accepted methods. Notwithstanding, Mobile device forensics is considered an evolving specialty in the field of digital forensics (24,26). The procedures for the validation, preservation, acquisition, examination, analysis, and reporting of digital information had been discussed (26). Although there had been established programme and guidelines, mobile device forensics, like any other evolving field still has its own forensics challenges (27). Evolvement on different areas of usage for mobile phones provides even further challenges in their investigation. I.e. investigation of cloud applications on mobile phones (45,47–49), malwares on smartphones (50–52), and investigating mobile phones as part of botnets (53) and SCADA (54) systems are all challenging forensics research areas. In view of the evolving nature of mobile device forensics, it is suggested that forensic practitioners who rely primarily on general-purpose mobile forensic toolkits might find that no single forensic tool could recover all relevant evidence data from a device (6). Therefore, researchers are working to establish the best forensic tools and procedures that are reliable for mobile device's investigation (28,32,34,47,55).

Investigations conducted in the field of mobile device forensics still show variations in research opinions on the effectiveness and reliability of different forensic tools when applied to different mobile device architectures (24). The removal of internal memories from mobile device or their mirroring procedure is evasive and complex because of difficulties in having direct hardware access (16). To resolve such challenges, five mobile device forensic scenarios were studied; and a method to perform data acquisition of Android Smartphone regardless of the architecture was proposed. The method was validated using Motorola Milestone II A953, Sony Ericson Xperia X10 miniPro, Motorola Defy, Samsung Galaxy S 9000a, Motorola II, and Motorola Milestone A853 (16). These architecture-based difficulties were confirmed by investigating Symbian and Windows based mobile devices (24); which revealed that tools for forensic investigation of Smartphone mobile devices always would pose challenges in forensic investigation because of the continual evolving nature of the technology (24,41). The data recovery capabilities of EnCase, FTK, Recuva, R-Studio and Stellar Phoenix from a desktop Windows XP were compared (37). The comparison revealed that EnCase, FTK, Recuva and R-Studio performed identically when recovering marker files from most images (37,54). The experiment showed also that Stellar Phoenix corrupted two bytes in each of two text files (even when these files had not been deleted) and added a padding of zeroes at the end of another file that had not been deleted (37). Nonetheless, the study concluded that no two tools produced identical results (37). Further to this, a different study conducted on Samsung Star 3G phone used TK file Explorer 2.2, MOBILedit 4 and Samsung PC studio for logical acquisition of different data files for analysis, and this was to create a framework for forensic investigation of Samsung Star 3G device (17). Evidence collection and analysis conducted on Nexus 4 phone discovered a flaw that allowed access to all data on the device without device wipe that occurs when the bootloader is unlocked (56). The challenges of smartphone forensics continue to expand even with the emergence of Linux-based Firefox OS, which has no procedure yet for forensic investigation (57).

To ensure a sound forensic investigation, care should be taken to preserve and retrieve the volatile data inside the memory of the mobile device (48); hence a backup and acquisition process was proposed to work on windows mobile phones, android mobile phones and iPhones (48). Recovery of information held on Windows Mobile smart phone was investigated using different approaches to acquisition and decoding, excepting AccessData FTK and DD imagers. The investigation concluded that no one technique recovers all information of potential forensic interest from the device (36). Further work on mobile device forensics show that a diverse collection of smartphone forensic tools has been introduced; however, these studies do not guarantee data integrity, which is required for digital forensic investigation (49). Therefore, Android device acquisition utilizing Recovery Mode was investigated to analyse the Android device Recovery Mode variables that compromise data integrity at the time of acquisition. Consequently, an Android data acquisition tool that ensures integrity of acquired data was developed (49). It was noted there was not yet specific procedures or rules to collect evidence from Smartphone (7). Therefore, forensic investigators use existing procedures in the acquisition of digital evidences (7). Furthermore, it was suggested that the relative amount of important evidences that could be gathered from the Smartphone differ based on different versions of software system that runs on the Smartphone. However, nothing was done on the acquisition of evidence on a formatted android device - where it was claimed that all data and applications were erased (7).

The challenges of data recovery for forensic investigation extends to cloud computing environment (47). Therefore, utilizing TPM in hypervisor (43), implementing multi-factor authentication and updating the cloud service provider policy to provide persistent storage devices are proposed to overcome the difficulties in Cloud forensics (52,58); which include limited access to obtain evidence from the cloud, seizure of physical evidence for integrity validation, evidence presentation, or rulings on data saved in different locations (47). In a cloud-related forensic investigation, Guidance Encase and AccessData forensic Toolkit were evaluated; and the tools show that they can successfully return volatile and non-volatile data from the cloud (51). Thus, a foundation is laid for the development of new acquisition methods for the cloud that will be trustworthy and forensically sound (51). To ensure a reliable cloud-based forensic investigation, a step-by-step technique for evidence data collection was proposed (56). There was a review of seven years of research into forensic investigation of various smartphone mobile device platforms, data acquisition scheme and information recovery methods in order to provide a comprehensive reference material to enhance future research (34). Prior to the advancement of forensic tools, the traditional method of memory acquisition focused on the physical memory. This procedure most often requires the removal of the memory chip from the chipboard. These methods put valuable evidences at risk because during the removal process there might be loss or damage of essential evidences.

Emphasizing on the need for accurate and reliable forensic tools and procedures, there was a warning against unfair application of wrong forensic techniques and evidence to secure conviction in fervour of the prosecution (53). Notably, some challenging areas include "erroneous allegations of knowledgeable possession, misuse of time stamps and metadata, control and observation of the discovery process". Other challenging areas include "authentication issues, deficiencies and the lack of verification for proprietary software tools, deliberate omission or obfuscation of exculpatory evidence and inadvertent risks resulting from the use of legitimate services" (53).

The foregoing review shows there is need to investigate the recovery performance of Phone Image Carver, AccessData FTK, Foremost, Recover My Files, and DiskDigger from FTK and DD images acquired from Android Mobile Smartphones.

## 3.0 EXPERIMENT SETUP

In our experiments, we used the popular Samsung Galaxy S2 i9100 as the case study device. The device has an internal memory of 16GB, Random Access Memory (RAM) of 1GB, running Android Gingerbread version 2.3.4 Operating System (OS) (Android OS, Ice Cream Sandwich version 4.0.3). The focus of our study was on the internal memory; thus, no memory card was inserted in the phone. Prior to the experiments, the phone was preloaded with Enron dataset (59–65), which is considered similar to data collected for fraud detection. Therefore, it is a good dataset for the present investigation. The device was subsequently factory-reset before taking images of the phone (see Figure 1). The reason for factory resetting the phone was to wipe all preloaded data and investigate the effectiveness of the forensic tools in recovering the data erased from the device.

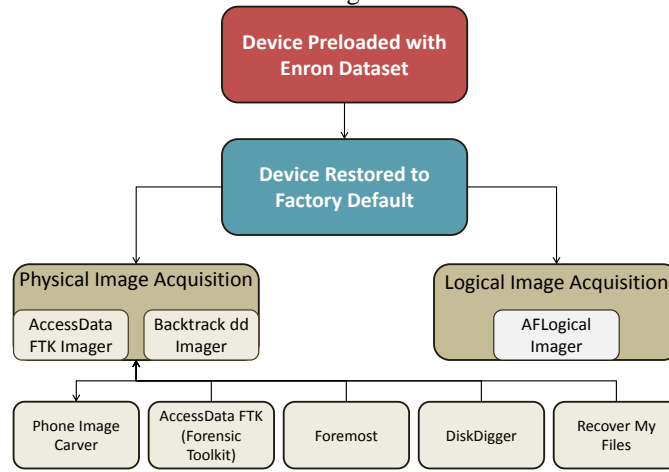

**Figure 1: Schematic representation of processes conducted on the case study device**

Figure 1 Shows that two different image acquisition processes, Logical and Physical image acquisition, were conducted after the device was restored to default factory state. AFLogical forensic tool was used for logical Image acquisition on the formatted disk (device). The tool captures the Call-Log calls, Contacts Phones, MMS, MMS-Parts and SMS, which were contained in the preloaded Enron dataset. It stores this information in a zip folder named forensics.zip within the device itself. The .zip folder contains .csv files, which hold the logs of the device. However, when opened .cvs files show blank suggesting that logical acquisition of data is not executable in a reformatted device. Comparing to another investigation, analysis of logical image, bb file, from Blackberry PlayBook device did not produce direct data files of user activity. However, it produced some key files that can assist to further trace device usage (34,54,66). Related studies confirm also that Encase and FTK forensic tools could not recover all data from NTFS-formatted logical disk partitions (35,53).

On the other hand, AccessData FTK Imager 3.1.3.2 and Backtrack dd Imager were used for Physical acquisition of images. Subsequently, Phone Image (Carver v1.6.0), AccessData FTK (Forensic ToolKit), Foremost, DiskDigger and Recover My Files (v4.7.2) were employed to analyse the two different Physical Images.

During the image creation process using AccessData FTK Imager 3.1.3.2, the backup option was not selected. This was to ensure that no backup data was available. The physical memory of the device was imaged and analysed using several tools. The resulting image revealed that the tool created only 2.227GB image file compared to the 16GB physical memory capacity of the device. This means that AccessData FTK imager recovered only 14% of the device memory capacity. The imager separated the physical drive into 8 images. The size of the first seven images is 1.46GB and the eighth image is 767MB. This experiment compares with another study where AccessData FTK Imager recovered a higher average of 86.4% of the physical disk capacity from the various image segments (35).

Another image was acquired from the case study device using .dd imager in Backtrack. This tool converts the memory into a disk dump. Similar to the acquisition of image using AccessData, the dd imager used command line prompt in Backtrack dd Imager without any filter. The image size converted by this tool shows only 11GB of the 16GB memory capacity of the device, which translates to 68.75% recovery of the device capacity. The remaining memory was ignored due to slack spaces, which is the disk space between the end of the file content and the end of the last cluster in which the file is saved (35). Two types of disk dumps were created after imaging the device. One of the disk dumps uses the access data FTK images, while the other uses backtrack dd image. Subsequently, several recovery tools were used to perform analysis on the images to determine the effectiveness of these tools. Prior to analysis, validation checks were conducted on the images using Hash Calculation for integrity checking (49) as shown in Table 1. Image numbers S2.001 – S2.008 in column one of Table 1 are the validation and integrity check identification numbers for each of the eight images acquired by AccessData FTK imager. DD, on the other hand is the validation and integrity check ID number for the DD image that was acquired as a single image. The corresponding MD5 sums and SHA1 values are given in columns two and three of Table 1.

**Table 1: Validation and Integrity Check of Images**

| Image | MD5 | SHA1 |
|---|---|---|
| S2.001 | df14a97ed884e959f79d718a4a3e8de0 | 838f1291740988d86e2b2b22625646e20e9e535a |
| S2.002 | 334bc971671ad78a09abadd81aa2419f | aa42e0269f4cb2fd53b065f0aea56823fb770d88 |
| S2.003 | 53393c41f197b08a693db24600b2eab1 | 35cc22977e11e6c966c569260812fde04471401e |
| S2.004 | 4396a40fb1825166db005e39d211b5a8 | fecb8eabff34ea01ad4a84cc283625af5ca0319f |
| S2.005 | 6e9a8fc2cc1235da1d33a51d73a53c30 | 1b03158408ae5e567d6a7e27993ab46ecd8fe686 |
| S2.006 | Ab49d0350af243d6d3d6df1791adb58f | Eb7a722cc1b14fc8bdf06c81390df45ab34f5a50 |
| S2.007 | 35d8479c1dc29edacb33ad9dcaa07d5b | Eac0ba1a3e4e8ef8b0195ab682081c4d12d9d36e |
| S2.008 | 01a5c72710d2223d012e8e7b71e9055c | C782fb92564990314de7baefa2db748ac186aa7b |
| DD | 1eeac023329e6d70ffcc78e7230c1ca7 | 76ae66d29894ae6b21f73bac87578c9dd1202c77 |

## 4.0 RESULTS AND DISCUSSIONS

Two types of image acquisition namely logical image and Physical image acquisition were conducted. AFLogical was used to acquire the logical disk image. The tool captures the Call-Log calls, Contacts Phones, MMS, MMS-Parts and SMS that were contained in the preloaded Enron dataset. It stores this information in a zip folder named forensics.zip within the device itself. The .zip folder contains .csv files that holds the logs of the device. However, when opened the .cvs files show blank suggesting that logical acquisition of data is not executable in a reformatted device. This revelation agrees with particularly the respective findings of Buchanan-Wollaston and Mercuri. The blank .cvs files confirm that Encase and FTK were unable to recover some data from NTFS-formatted logical disk partitions, except by further procedure where the data acquired must be decoded (34–36,53,67).

Earlier on AccessData image and backtrack dd image were acquired for the Physical disk of the device. The content of the FTK Image and BackTrack dd image of the target Samsung Mobile device were loaded on different forensic tools. Three different tools namely, Phone Image Carver v. 1.6.0, AccessData FTK (Forensic ToolKit 1.81.3), and dd Image FTK were used to analyse the images.

**Table 2: Recovered Files Format using Phone Image Carver**

| | File Format | | | | | | | | | |
|---|---|---|---|---|---|---|---|---|---|---|
| FTK Image | DocX | HTML | JPG | MP3 | SQLite | SWF Flash | Text | Text UTF-16 | Text-Shift-JIS | Zip |
| Backtrack dd Image | DocX | HTML | JPG | MP3 | SQLite | SWF Flash | Text | Text UTF-16 | Text-Shift-JIS | Zip |

Table 2 shows the various files format recovered from the images using Phone Image carver. Phone Image Carver recovered evidence of many different formats of data, confirming that Phone Image Carver supports over 300 file types (36). The present experiment supports the finding which suggests that .docx files were the only Office documents detected. However, it disagrees that Phone Image Carver did not detect .jpg files (36). This study reveals that utilizing Phone Image Carver tool is extremely time-consuming and not efficient; suggesting that it is not suitable for real investigation. Phone Image Carver tool does not list out the deleted files according to file formats, however the information still exists within the image. It was noted that Phone Image Carver does not permit addition of further file types to its database, meaning that a number of file types from the data set would not be detected (36). An additional feature of Phone Image Carver tool is that the activities performed in it are recorded in a log file. This offers a great opportunity for the analyst to go back and review his steps each time he uses Phone Image Carver tool.

The FTK and dd images acquired in Section 3.0 were analysed using AccessData FTK (Forensic Tools Kit). Analysis revealed that twelve (12) categories of files were recovered from each of the two Images as shown in Figure 2.

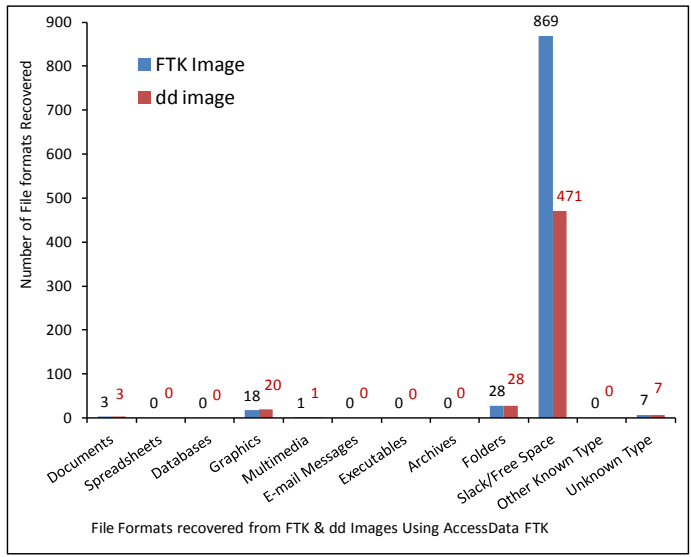

**Figure 2: Recovery and Analysis of Data from Both FTK and dd Images Using FTK Forensic Toolkit**

Figure 2 shows minor differences in the recovery function of both FTK and dd images from the same restored device. The recovered files in Figure 2 reveals that different images from different acquisition tools give different depth of evidence recovered in the analysis (35). While FTK Image recovered total file items of 926, dd Image recovered far less number of 530 when the same FTK tool was used on both images (Table 3 & Table 4). It is clear that entries in some of the file items are the same, some with zero entries (no recovery); however, few others show significant differences between both images. Worthy of note are the entries under FTK Image for "Unchecked Items (926), Other Thumbnails (18) and Filtered In (926)."

**Table 3: FTK Image result using FTK**

| File Items | No. | File Status | No | File Category | No |
|---|---|---|---|---|---|
| Total File Items | 926 | KFF Alert Files | 0 | Documents | 3 |
| Checked Items | 0 | Bookmarked Items | 0 | Spreadsheets | 0 |
| Unchecked Items | 926 | Bad Extension | 18 | Databases | 0 |
| Flagged Thumbnails | 0 | Encrypted Files | 0 | Graphics | 18 |
| Other Thumbnails | 18 | From E-mail | 0 | Multimedia | 1 |
| Filtered In | 926 | Deleted Files | 2 | E-mail Messages | 0 |
| Filtered Out | 0 | From Recycle Bin | 0 | Executables | 0 |
| | | Duplicate Items | 2 | Archives | 0 |
| | | OLE Sub-items | 0 | Folders | 28 |
| | | Flagged Ignore | 0 | Slack/Free Space | 869 |
| | | KFF Ignorable | 0 | Other Known Type | 0 |
| | | Data Carved Files | 0 | Unknown Type | 7 |

**Table 4: dd Image Result using FTK**

| File Items | No. | File Status | No | File Category | No |
|---|---|---|---|---|---|
| Total File Items | 530 | KFF Alert Files | 0 | Documents | 3 |
| Checked Items | 0 | Bookmarked Items | 0 | Spreadsheets | 0 |
| Unchecked Items | 530 | Bad Extension | 20 | Databases | 0 |
| Flagged Thumbnails | 0 | Encrypted Files | 0 | Graphics | 20 |
| Other Thumbnails | 20 | From E-mail | 0 | Multimedia | 1 |
| Filtered In | 530 | Deleted Files | 2 | E-mail Messages | 0 |
| Filtered Out | 0 | From Recycle Bin | 0 | Executables | 0 |
| | | Duplicate Items | 2 | Archives | 0 |
| | | OLE Subitems | 0 | Folders | 28 |
| | | Flagged Ignore | 0 | Slack/Free Space | 471 |
| | | KFF Ignorable | 0 | Other Known Type | 0 |
| | | Data Carved Files | 0 | Unknown Type | 7 |

The analysis shows the corresponding entries for dd Image are lower, except for graphics and other thumbnails where dd Image and FTK Image recovered virtually equal numbers of files. This suggests that dd images provide negligently better recovery for graphic and thumbnail files. However, both images recorded similar values under File Status and File Category, except for Slack/Free Space where FTK Image had significantly higher value (869) than dd Image (471). The analysis suggests that in the event of recoverable evidences from slack spaces, FTK image gives more recovery files than dd image. In other words, slack spaces could contain deleted files, deleted file fragments and hidden data (55,57). The negative side of it is that recovering from it could result in a waste of time if there were no recoverable evidence in the slack spaces. It is worthy to note that files were not recovered from FTK Image 8, which contains the Slack/Free Spaces. On the percentage file recovery of FTK Image, Mercuri noted that CFTT Program tests revealed defects in the hard disk image preparation process on Windows XP OS (35,53). Taken into account, the analysis revealed that defect might still have some drawback on the performance of FTK as an imager in the present study on smartphone.

Further analysis was conducted on the FTK and Backtrack dd Images using Foremost in Backtrack forensic tool. The FTK Images (1-8) were analysed in segments, the way they were imaged. Figure 3 shows the numbers of various file formats recovered from the two images.

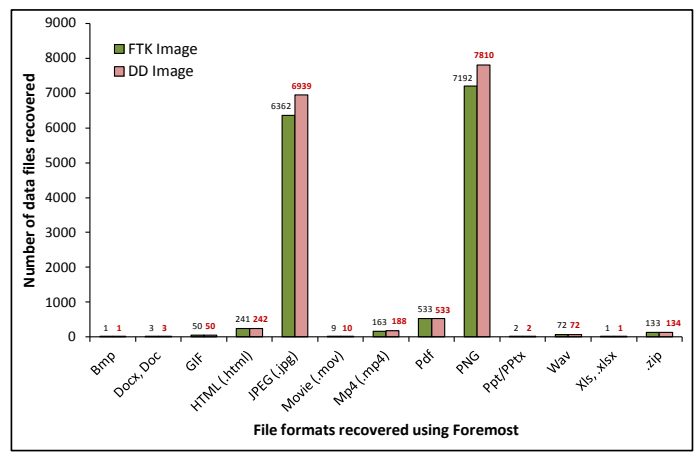

**Figure 3: Number of Data files recovered from FTK and dd Images Using Foremost in Backtrack**

Unlike AccessData FTK, analysis conducted using Foremost showed format-specific files from both Images. Apparently, Foremost recovered similar types of files from the two images, particularly, bmp, docx/doc, gif, html, movie, pdf, ppt/pptx, wav, xls/xlsx and zip. The analysis showed that both images recovered the largest number of jpeg and png files. While analysis in Foremost showed that FTK image recovered 6362 jpeg and 7192 png files, dd image on the other hand recovered 6939 and 7810 respectively. The large number of graphics analysed (recovered) suggests that the slack spaces contain more deleted graphic files, therefore foremost performed better in recovering them (55). However, analysis in Foremost revealed a higher number of files from dd Image than it did from FTK image for jpeg/jpg, mp4, and png. Compared to the AccessData FTK (Forensic Tools Kit), Foremost analysed a higher number of file types. The recovery of high number of file types could be a result of high-performance of Foremost tool. The absence of slack/free spaces in the analysis suggests that Foremost performed better in recovering all deleted files, deleted file fragments and hidden data files that were present in the slack/free spaces of the images, which AccessData FTK could not recover as shown in Figure 2 (55,57).

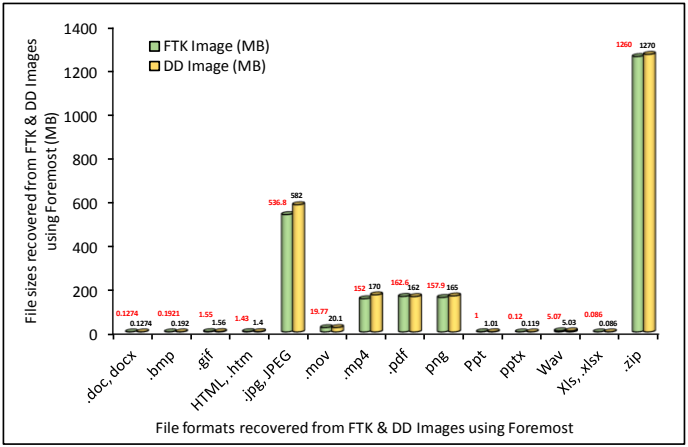

**Figure 4: Data file sizes recovered from FTK and dd Images Using Foremost in Backtrack**

The large sizes of data file shown in Figure 4 further corroborates the suggestion that Foremost recovered deleted files, deleted files fragments and perhaps hidden files that might be contained in the slack/free spaces which AccessData FTK could not recover as shown in Figure 2 (55). Comparing the data file sizes with the number of files recovered in Figure 3, it is evident that Foremost forensic tool recovered more data from Backtrack dd Image than it recovered from FTK Image. This suggests also that Foremost forensic tool performs better in analysing data files from Backtrack dd Image than FTK Image. The analysis reveals also that the zip, jpeg, mp4, png, and pdf files were the files most recovered using Foremost forensic tool. This result shows that Foremost performs better than AccessData FTK in recovering files from FTK and dd images, particularly dd image of a restored Android Mobile Smartphone device.

The next forensic tool to use in the analysis of the two images (FTK and dd acquired images) was Recover My Files version 4.2.7. The result of the analysis is shown in Figure 5. The number of data files recovered from the two images shows no differences in the performance of Recover My Files as a forensic tool. The tool recovered 16,756 files from the FTK image, which is comparable with 16,758 files recovered from Backtrack dd image.

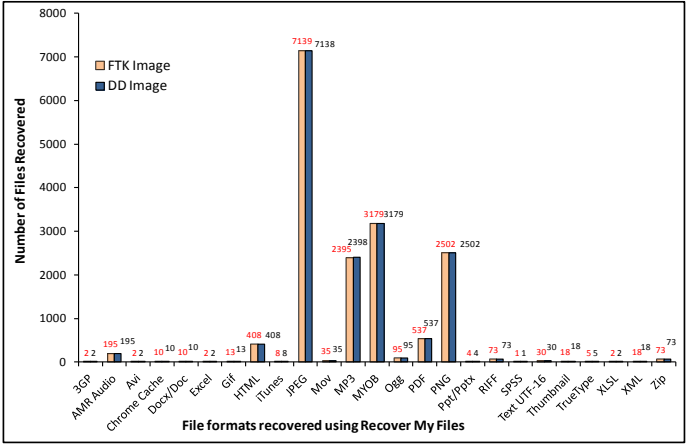

**Figure 5: Number of data files formats recovered from FTK and dd Images Using Recover My Files**

The results show also that jpeg, myob, png, mp3, pdf, html and amr audio files were the most recovered in the order of listing. The tool has shown to recover more file formats and showed no slack/free space. The additional file formats it recovered include 3gp, amr audio, avi, chrome cache, itunes, mp3, myob, ogg, tif, spss, text UTF-16, thumbnails, and truetype. Similarity in the recovery function of Recover My Files on the two images is revealed further in Figure 6. It is evident in figure 6 that the tool recovered equal sizes of data files from both FTK and dd images. The tool recovered 7.37GB of data files for each file image confirming that mov, zip, jpeg, myob, mp3, avi, png, and pdf are among the most recovered files format.

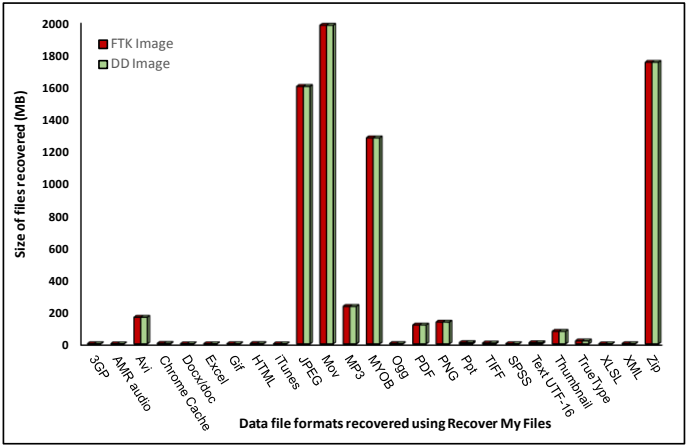

**Figure 6: Sizes of data files recovered from FTK and dd Images Using Recover My Files**

Comparing Recover My File and Foremost in Figure 7 and Figure 8, the results show that Foremost recovered 14,762 and 15,985 data files from FTK and dd images respectively while Recover My File recovered 16,756 and 16,758 data files respectively from the same images. This recovery performance is repeated in the size of data files recovered, where Foremost recovered 2.35GB and 2.38GB of data files from FTK and dd images. On the other hand, Recover My File recovered 7.37GB and 7.39GB data files respectively from the same FTK and dd images.

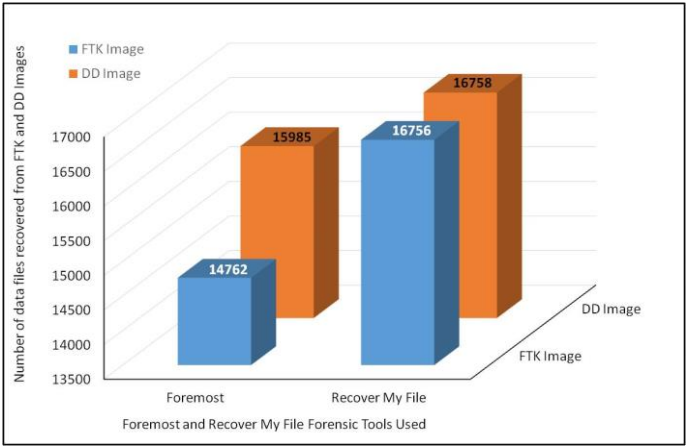

**Figure 7: Number of files recovered between Foremost and Recover My File**

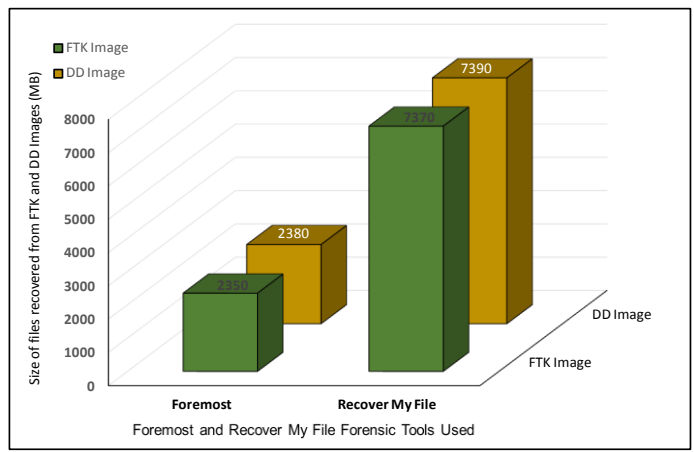

**Figure 8: Size of files recovered between Foremost and Recover My File**

Both the number and size of data files recovered could elaborate the huge difference in the recovery functions of the two forensic tools. While results show that foremost recovered large number of data files formats in jpeg (6939), png (7810), pdf (533) and hmtl (242) for FTK and dd Images, Recover my File recovered an average of jpeg (7139), myob (3179), png (2502), mp3 (2398), pdf (537) and html (408). Similarly, Foremost recovered an average size in jpeg (0.582GB), zip (1.27GB), mp4 (0.17GB), pdf (0.162GB) and png (0.165), while Recover My File recovered some phenomenal sizes in mov (1.98GB), zip (1.75GB), jpeg (1.6GB), myob (1.28GB), mp3 (0.2329GB), and avi (0.1651GB) for the two images. These analyses reveal that Recover My File is more effective than Foremost in recovering data files from FTK and dd images of a restored device.

The superior performance of Recover My File over Phone Carver image, AccessData FTK and Foremost was exciting. Therefore, one more analysis was conducted on the acquired FTK and Backtrack dd Images using DiskDigger forensic tool. The analyses are shown in Figure 9.

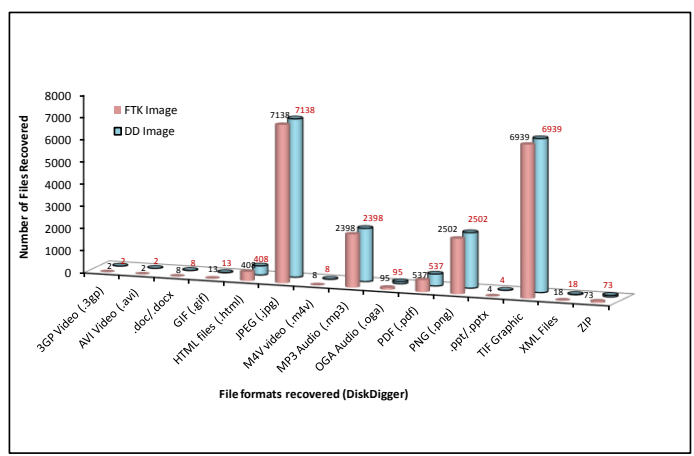

**Figure 9: FTK and DD Image Analysis of Number of Data Files Using DiskDigger Tool**

Figure 9 reveals that DiskDigger recovered also many different data files format. What is interesting here is that DiskDigger recovered equal number of data files from both FTK and DD Images. The figure shows that the largest numbers of files recovered from both images are from jpeg (7,178), tif (6939), png (2502), and mp3 audio (2398) files in the listed order. It is important to observe that DiskDigger tool did not record any slack/free space. This implies that the tool made deep recovery of all files in the two images. The distribution of the number of recovered data files [see Figure 9] corresponds to the sizes of data files recovered from the images [see Figure 10].

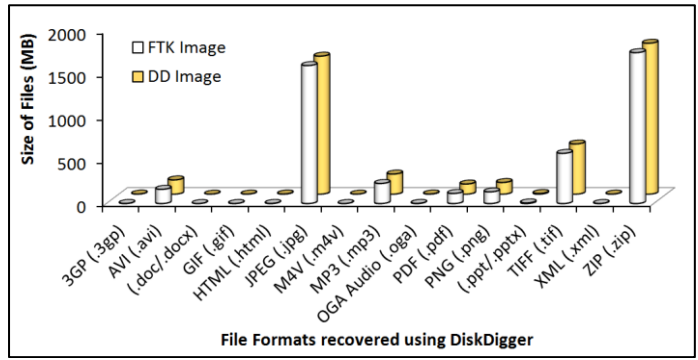

**Figure 10: FTK and DD Image Analysis of Size of Data Files Using DiskDigger Tool**

As shown in Figure 10, the zip folder recorded the largest data file sizes, while jpeg, tif, mp3, avi, and mp3 followed in the same order.

The results in Figure 10 suggest that Recover My File and DiskDigger show better recovery performance than Phone Image Carver and AccessData FTK. Although DiskDigger recovered 15 different file formats, Recover My File recovered 25 file formats, indicating a better deep recovery performance. However, DiskDigger recovered a greater total number of files (20145) for both Images than the total number of files (16758) recovered by Recover My File as shown in Figure 11. The greater number of data files recovered from DiskDigger comes from the number of .tif files (6939) for both FTK and dd Images, which Recover My File did not recover.

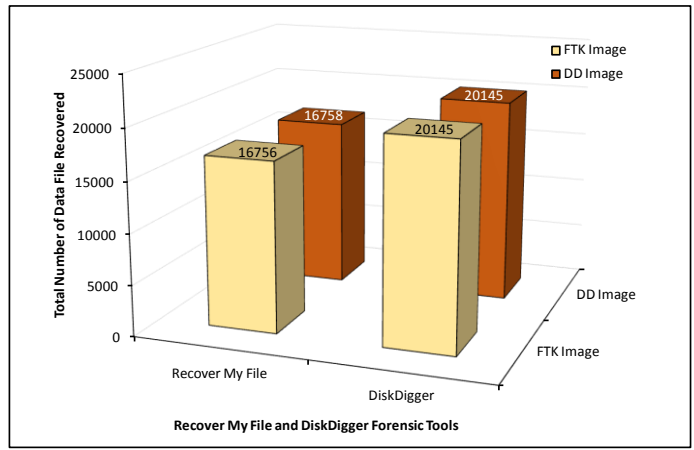

**Figure 11: Comparing Data File Number Recovered Using Recover My File and DiskDigger**

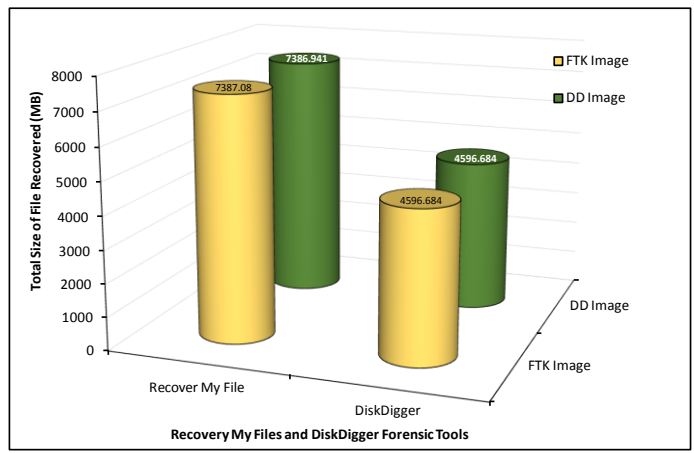

**Figure 12: Comparing Data File Size Recovered Using Recover My File and DiskDigger**

Comparing Recover My File and DiskDigger in terms of data file size, it is clear that Recover My File recovered a larger file size (7.387GB) than the data file size (4.596GB) recovered by DiskDigger shown in Figure 12. This difference is explained by the larger number of file formats recovered by Recover My File. These additional file formats and sizes (MB) include, amr audio (0.496MB), chrome cache (2.6MB), itunes (0.121MB), mov (1980MB), myob (1280MB), ogg (1.8MB), spss (0.657MB), text UTF-16 (7.3MB), thumbnails (77.7MB), and truetype (18.3MB). It is evident that the two additional data file formats from Recover My File that made major contribution to the difference are .mov and .myob. The results suggest that DiskDigger performs better than Recover My File only in the area of the number of data files, particularly the .tif files (6939), recovered by each from the two images. Conversely, Recover My File performs better in terms of deep search for different data file formats, and the size of data files recovered.

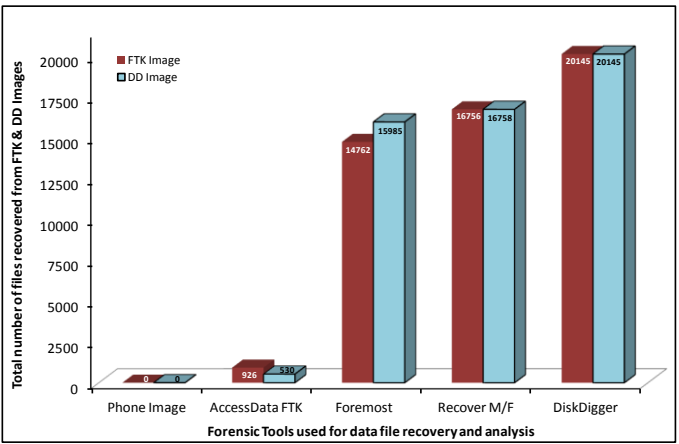

Figure 13: Total number of files recovered from FTK & DD Images using different forensic tools

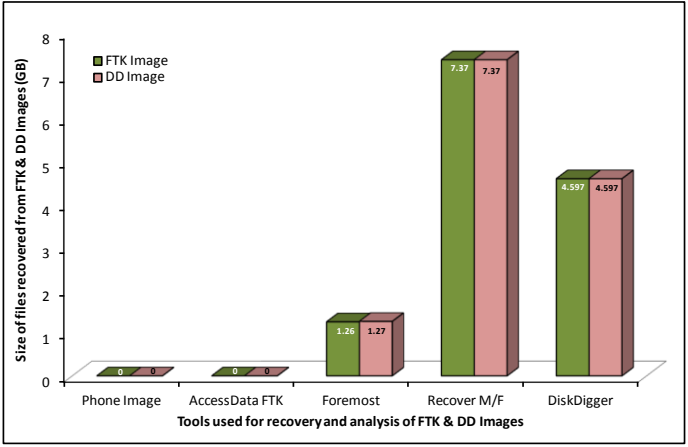

Figure 14: Total size of data files recovered from FTK & DD Images using different forensic tools

The summary of the recovery performance of the five forensic tools is shown in Figure 13 and Figure 14. The figures show that Phone Image Carver AccessData FTK did not perform well under the present experimental conditions as a forensic recovery tool. Foremost, on the hand recovered more file formats and appreciable large number of data files with corresponding data file size. However, Foremost shows to recover slightly higher number of data files (jpeg and png) from dd image than FTK image while the sizes of the data file recovered show no difference. DiskDigger appears to have performed well compared to Recover My File. It recovered many file formats that is still less than the number of data file formats recovered by Recover My File. DiskDigger proves to recover the highest number of data files but less than the size of data files recovered by Recover My File. Therefore, the study shows that Recover My File has the best recovery function as a forensic tool. It proved to have the deepest search penetration, recovered the highest number of data file formats, and recovered the largest size of data file formats, although it was less than DiskDigger in the number of data files recovered.

## 5.0 CONCLUSION & FUTURE WORKS

The increasing use of smartphone for various social-economic transactions led to a consequent increase in cybercrimes committed through smartphones. The nature of these devices and the variety of applications resided on them deemed challenging challenges to forensic investigators. To address these challenges, this paper investigated the use of different forensic tools for recovering data files from a restored Android mobile phone. The data was extracted using different forensic tools, namely AccessData FTK and Backtrack dd, on the physical image acquisition of the device. The focus of this paper was to investigate the data recovery functions of Photo Image Carver, AccessData FTK, Foremost, Recover My Files and DiskDigger in forensic investigation of FTK and DD images from Smartphone mobile device. The study revealed that .dd images compare more favourably for android mobile forensic investigation than FTK images, judging from the size of evidence it holds. Moreover, the experiment revealed that Phone image carver recovered some file types including SQLite, SWF Flash, and Text-Shift JIS, which other tools could not recover. Phone Image Carver keeps log file record of activities performed in it, giving the analyst the opportunity to review his previous steps each time he uses the tool. Foremost proves to recover more number of files data than Phone Image Carver and AccessData FTK. On the other hand, Recover My Files and DiskDigger had greater percentage recovery performance than Foremost, Phone Image Carver, and AccessData FTK. Both Recover My Files and DiskDigger recovered many data file formats suggesting they had a deep penetration recovery capability. However, Recover My Files recovered more files of type mov, zip, JPEG, and MYOB than DiskDigger recovered. Recover My Files proves to recover the greatest percentage of evidence by recovering 3GP, AMR audio, avi, itunes, Myob, ogg, thumbnails and truetype files, which no other tools recovered. Therefore, Recover My Files proves to be the best recovery tool in this study.

In conclusion, the analysis tools used in this experiment showed different levels of recovery performance. Most of the tools recovered major file formats that other tools did not recover, suggesting that no single forensic tool could recover all forensic evidences in a smartphone image. In future, further similar studies are suggested to be conducted on other mobile platforms such as iPhone and compare and contrast results with those presented in this paper.